\documentclass[12pt,epsf,epsfig]{article}
\usepackage{amsmath}

\begin{document}

\begin{center}
{\huge{Galileon string measure and other modified measure extended objects.}} \\
\end{center}

\begin{center}
T.O. Vulfs \textsuperscript{1,2,3}, E.I. Guendelman \textsuperscript{1,3} \\
\end{center}

\begin{center}
\textsuperscript{1} Department of Physics, Ben-Gurion University of the Negev, Beer-Sheva, Israel \\
\end{center}

\begin{center}
\textsuperscript{2} Fachbereich Physik der Johann Wolfgang Goethe Universit\"{a}t, Campus Riedberg, Frankfurt am Main, Germany \\
\end{center}

\begin{center}
\textsuperscript{3} Frankfurt Institute for Advanced Studies, Giersch Science Center, Campus Riedberg, Frankfurt am Main, Germany \\
\end{center}

E-mail: vulfs@post.bgu.ac.il, guendel@bgu.ac.il

\abstract

We show that it is possible to formulate string theory as a "Galileon string theory". The galileon field $\chi$ enters in the definition of the integration measure in the action. Following the methods of the modified measure string theory, we find that the final equations are again those of Polyakov. Moreover, the string tension appears again as an additional dynamical degree of freedom. At the same time the theory satisfies all requirements of the galileon higher derivative theory at the action level while the equations of motion are still of the second order. A galileon symmetry is displayed explicitly in the conformal string worldsheet frame. Also we define the galileon gauge transformations. Generalizations to branes with other modified measures are discussed.

\section{Introduction}

General Relativity has been the leading theory of gravity for many years. But there are several problems mainly concerning the dark sector that couldn't be explained in that framework. So several ideas of modifying GR are now circulating. One of them is the "Galileon gravity" \cite{i,h,j}. It is a specific scalar-tensor theory that comes from the braneworld paradigm.

On the other hand String Theory is a candidate for the theory of all matter and interactions including gravity. But it has a dimensionfull parameter, the tension of the string, in its standard formulation.

Previously, in the framework of Two Measure Theory, originally formulated as a theory of gravity \cite{d,b}, the tension was derived as an additional degree of freedom \cite{a,c,e}.

The main idea of this paper is that the modified by the additional scalar field - the galileon - measure leads to the bosonic string action without an ad hoc dimensionfull scale. The galileon symmetry is also well defined in the conformally flat worldsheet gauge. Furthermore, this formulation realizes the goal of the galileon introducing a higher derivative action while the equations of motion remain second order only.

This paper is organized as follows. In Section 2 we review the two-measure theory in the string context. In Section 3 a new measure of integration which possesses a galileon symmetry is introduced. In Section 4 we study a galileon strings. A discussion of generalizations to branes is given in Section 5. In Section 6 the galileon gauge transformations are discussed. In Section 7 we provide the results and conclusions. \\

\section{Two-Measure Theory}

The standard Polyakov-type action for the string:

\begin{equation} \label{eq:1}
S_{Polyakov} = -T\int d^2 \sigma \frac12 \sqrt{-\gamma} \gamma^{ab} \partial_a X^{\mu} \partial_b X^{\nu} g_{\mu \nu}.
\end{equation}

Here $\gamma^{ab}$ is the intrinsic Riemannian metric on the 2-dimensional string worldsheet and $\gamma = det(\gamma_{ab})$; $g_{\mu \nu}$ denotes the Riemannian metric on the emdedding spacetime. $T$ is a string tension, a dimensionfull scale introduced into the theory by hand. \\

From the variations of the action with respect to $\gamma^{ab}$ and $X^{\mu}$ we get the following equations of motion:

\begin{equation} \label{eq:tab}
T_{ab} = (\partial_a X^{\mu} \partial_b X^{\nu} - \frac12 \gamma_{ab}\gamma^{cd}\partial_cX^{\mu}\partial_dX^{\nu}) g_{\mu\nu}=0,
\end{equation}

\begin{equation} \label{eq:3}
\frac{1}{\sqrt{-\gamma}}\partial_a(\sqrt{-\gamma} \gamma^{ab}\partial_b X^{\mu}) + \gamma^{ab} \partial_a X^{\nu} \partial_b X^{\lambda}\Gamma^{\mu}_{\nu\lambda}=0,
\end{equation}

where $\Gamma^{\mu}_{\nu\lambda}$ is the affine connection for the external metric. \\

There are no limitations on employing any other measure of integration different than $\sqrt{-\gamma}$. The only restriction is that it must be a density under arbitrary diffeomorphisms (reparametrizations) on the underlying spacetime manifold. The two-measure theory is an example of such a theory. The title indicates that we could use the standard and the modified measure simultaneously. The action with a modified measure only is quite sufficient for our purpose. \\

In the framework of this theory two additional worldsheet scalar fields $\varphi^i (i=1,2)$ are introduced. A new measure density is

\begin{equation}
\Phi(\varphi) = \frac12 \epsilon_{ij}\epsilon^{ab} \partial_a \varphi^i \partial_b \varphi^j.
\end{equation}

Then the modified bosonic string action is

\begin{equation} \label{eq:5}
S = -\int d^2 \sigma \Phi(\varphi)(\frac12 \gamma^{ab} \partial_a X^{\mu} \partial_b X^{\nu} g_{\mu\nu} - \frac{\epsilon^{ab}}{2\sqrt{-\gamma}}F_{ab}(A)),
\end{equation}

where $F_{ab}$ is the field-strength  of an auxiliary Abelian gauge field $A_a$: $F_{ab} = \partial_a A_b - \partial_b A_a$. \\

To check that the new action is consistent with the Polyakov one, let us derive the equations of motion of the action (\ref{eq:5}). \\

The variation with respect to $\varphi^i$ leads to the following equations of motion:

\begin{equation} \label{eq:6}
\epsilon^{ab} \partial_b \varphi^i \partial_a (\gamma^{cd} \partial_c X^{\mu} \partial_d X^{\nu} g_{\mu\nu} - \frac{\epsilon^{cd}}{\sqrt{-\gamma}}F_{cd}) = 0.
\end{equation}

It implies

\begin{equation} \label{eq:a}
\gamma^{cd} \partial_c X^{\mu} \partial_d X^{\nu} g_{\mu\nu} - \frac{\epsilon^{cd}}{\sqrt{-\gamma}}F_{cd} = M = const.
\end{equation}

The equations of motion with respect to $\gamma^{ab}$ are

\begin{equation} \label{eq:8}
T_{ab} = \partial_a X^{\mu} \partial_b X^{\nu} g_{\mu\nu} - \frac12 \gamma_{ab} \frac{\epsilon^{cd}}{\sqrt{-\gamma}}F_{cd}=0.
\end{equation}

We see that these equations are the same as in the standard Polyakov-type formulation (\ref{eq:tab}), (\ref{eq:3}). Namely, taking the trace of (\ref{eq:8}) we get that $M = 0$. By solving $\frac{\epsilon^{cd}}{\sqrt{-\gamma}}F_{cd}$ from (\ref{eq:a}) (with $M = 0$) we obtain (\ref{eq:tab}). \\

A most significant result is obtained by varying the action with respect to $A_a$:

\begin{equation}
\epsilon^{ab} \partial_b (\frac{\Phi(\varphi)}{\sqrt{-\gamma}}) = 0.
\end{equation}

Then by integrating and comparing it with the standard action it is seen that

\begin{equation}
\frac{\Phi(\varphi)}{\sqrt{-\gamma}} = T.
\end{equation}

That is how the tension is derived as a constant of integration opposite to the standard equation (\ref{eq:1}) where the tension is put ad hoc. The idea of modifying the measure of integration proved itself effective and profitable. \\

The variation with respect to $X^{\mu}$ leads to the second Polyakov-type equation (\ref{eq:3}). \\

\section{Galileon String Theory}

Let us now take the measure of integration in another realization. In this case it is constructed from a scalar field $\chi$. A new measure density is

\begin{equation}
\Phi (\chi) = \partial_h (\gamma^{hd} \sqrt{-\gamma} \partial_d \chi).
\end{equation}

The scalar field $\chi$ turns out to be the galileon one. Let's see it. The galileon theory must be invariant under a galileon shift symmetry:

\begin{equation}
\partial_{a}\chi \rightarrow \partial_{a}\chi +b_{a},
\end{equation}

\begin{equation}
\chi \rightarrow \chi +b_{a}\sigma^{a},
\end{equation}

where $b_a$ is the arbitrary displacement vector and $\sigma^a = (\tau, \sigma)$. \\

The metric $\gamma^{cd}$ in 2D can always be taken in a certain gauge to be conformally flat. Then in the conformal worldsheet gauge

\begin{equation}
\gamma^{cd} = \exp(-\Phi(\chi)) \eta^{cd}, \qquad \sqrt{-\gamma} = \sqrt{\exp(2\Phi(\chi))}
\end{equation}

we obtain

\begin{equation}
\gamma^{hd}\sqrt{-\gamma} = \eta^{hd}.
\end{equation}

Therefore we see

\begin{equation}
\Phi(\chi) = \partial_h (\eta^{hd} \partial_d \chi).
\end{equation}

So the galileon symmetry is confirmed in the conformally flat frame. Notice that the galileon symmetry is an exact symmetry then.  \\

We are going to study the equations of motion with this new galileon measure in the new resulting string theory in the next section. We could rewrite it in the form that will be useful later:

\begin{equation} \nonumber
\Phi(\chi) = \sqrt{-\gamma} \nabla_h (\gamma^{hd} \partial_d \chi)=  \gamma^{hd} \sqrt{-\gamma} \nabla_h (\partial_d \chi) =
\end{equation}

\begin{equation}
 = \gamma^{hd} \sqrt{-\gamma} (\partial_h \partial_d \chi - \Gamma^e_{hd} \partial_e \chi).
\end{equation}

\section{Bosonic Strings with the Galileon Measure}

The modified string action:

\begin{equation} \label{eq:17}
S = -\int d^2 \sigma L = -\int d^2 \sigma \bar{L} \Phi(\chi),
\end{equation}

where $L$ is the Lagrangian density and $\bar{L}$ is the modified Lagrangian density. The explicit form of $\bar{L}$ is not anyhow assumed right now. We are going to get the constraint on it. \\

The variation of the action (\ref{eq:17}) with respect to $\gamma_{ab}$ gives us the density of the energy-momentum tensor $T^{ab}$:

\begin{equation}
T^{ab} = \frac{-2}{\sqrt{-\gamma}} \frac{\delta L}{\delta \gamma_{ab}}.
\end{equation}

We obtain

\begin{equation} \nonumber
T^{ab} =\frac{-2}{\sqrt{-\gamma}} (\frac{\partial L}{\partial \gamma_{ab}} - \partial_c (\frac{\partial L}{\partial \gamma_{ab,c}}))=
\end{equation}

\begin{equation}
=\frac{-2}{\sqrt{-\gamma}} (\frac{\partial \bar{L}}{\partial \gamma_{ab}}\Phi + \bar{L}\frac{\partial \Phi}{\partial \gamma_{ab}} - \partial_c(\bar{L}\frac{\partial \Phi}{\partial \gamma_{ab, c}}) -\partial_c(\Phi \frac{\partial \bar{L}}{\partial \gamma_{ab,c}})).
\end{equation}

The only thing that we now assume about $\bar{L}$ is that it is independent from $\gamma_{ab, c}$ where $\gamma_{ab, c}$ is a derivative of the metric with respect to the coordinates. Then the last term here is equal to 0. Then

\begin{equation} \nonumber
T^{ab} =-\frac{2}{\sqrt{-\gamma}} (\frac{\partial \bar{L}}{\partial \gamma_{ab}}\Phi + \bar{L} \gamma^{hd} \nabla_h (\partial_d \chi)\frac{\partial \sqrt{-\gamma}}{\partial \gamma_{ab}} + \bar{L} \sqrt{-\gamma} \nabla_h (\partial_d \chi) \frac{\partial \gamma^{hd}}{\partial \gamma_{ab}}-
\end{equation}

\begin{equation} \nonumber
-\bar{L}\sqrt{-\gamma} \gamma^{hd}\partial_e \chi \frac{\partial \Gamma^e_{hd}}{\partial \gamma_{ab}} +\partial_c (\bar{L} \gamma^{hd} \sqrt{-\gamma}\partial_e \chi \frac{\partial \Gamma^e_{hd}}{\partial \gamma_{ab,c}})) =
\end{equation}

\begin{equation} \nonumber
=-\frac{2}{\sqrt{-\gamma}} (\frac{\partial \bar{L}}{\partial \gamma_{ab}}\Phi + \frac12 \sqrt{-\gamma} \gamma^{ab}\bar{L} \gamma^{hd} \nabla_h \partial_d \chi -
\end{equation}

\begin{equation} \nonumber
- \frac12 \bar{L} \sqrt{-\gamma} (\nabla^{a} \nabla^{b} + \nabla^{b}\nabla^{a}) \chi + \frac12 (\gamma^{ae}\Gamma^b_{hd} + \gamma^{be} \Gamma^a_{hd}) \bar{L} \sqrt{-\gamma} \gamma^{hd} \partial_e \chi -
\end{equation}

\begin{equation}
+ \partial_c (\frac14 \bar{L} \sqrt{-\gamma} \gamma^{hd} \partial_e \chi (\gamma^{ae}(\delta^b_h \delta^c_d + \delta^b_d \delta^c_h)+\gamma^{eb}(\delta^a_h \delta^c_d + \delta^a_d \delta^c_h)-\gamma^{ec}(\delta^a_d \delta^b_h + \delta^a_h \delta^b_d)))).
\end{equation}

We could go to any preferable frame without the loss of generality. In the Lorentz frame:

\begin{equation} \nonumber
T^{ab} = -\frac{2}{\sqrt{\gamma}}((\frac{\partial \bar{L}}{\partial \gamma_{ab}}\Phi +\frac12 \bar{L}\sqrt{-\gamma} \gamma^{hd} \partial_h \partial_d \chi) - \frac12 \bar{L} \sqrt{-\gamma} (\partial^a \partial^b + \partial^b \partial^a) \chi +
\end{equation}

\begin{equation} \nonumber
+\partial_c (\frac12 \bar{L} \sqrt{-\gamma} \partial_e \chi (\gamma^{bc} \gamma^{ae} + \gamma^{eb} \gamma^{ac} - \gamma^{ec} \gamma^{ab}))) =
\end{equation}

\begin{equation} \nonumber
= -\frac{2}{\sqrt{-\gamma}} ((\frac{\partial \bar{L}}{\partial \gamma_{ab}}\Phi +\frac12 \sqrt{-\gamma} \bar{L} \gamma^{hd} \partial_h \partial_d \chi) + \frac12\sqrt{-\gamma}\partial^a \chi \partial^b \bar{L} + \frac12\sqrt{-\gamma} \partial^b \chi \partial^a \bar{L} -
\end{equation}

\begin{equation} \nonumber
- \frac12 \bar{L} \sqrt{-\gamma} \gamma^{ab} \gamma^{hd} \partial_h \partial_d \chi - \frac12 \sqrt{-\gamma} \gamma^{ab} \partial_e \chi \partial^e \bar{L}) =
\end{equation}

\begin{equation} \label{eq:22}
=-\frac{2}{\sqrt{-\gamma}} \frac{\partial \bar{L}}{\partial \gamma_{ab}}\Phi -\partial^a \chi \partial^b \bar{L} - \partial^b \chi \partial^a \bar{L} + \gamma^{ab} \partial_e \chi \partial^e \bar{L}.
\end{equation}

The trace equation:

\begin{equation}
\gamma_{ab}T^{ab} = -\frac{2}{\sqrt{-\gamma}} \gamma_{ab} \frac{\delta L}{\delta \gamma_{ab}} = 0.
\end{equation}

When taking a trace, we see that the last three terms in (\ref{eq:22}) are equal to 0. By the fact that $\frac{\partial \bar{L}}{\partial \gamma_{ab}} \gamma_{ab} = -\bar{L}$ we obtain that

\begin{equation}
\gamma_{ab}T^{ab} = \frac{2}{\sqrt{-\gamma}} \bar{L} \Phi(\chi) = 0.
\end{equation}

Excluding the trivial case $\Phi(\chi) = 0$, we get $\bar{L}=0$. That means that the higher-derivative terms in (\ref{eq:22}) are canceled. We are left with the usual modified string theory:

\begin{equation}
\frac{\delta L}{\delta \gamma_{ab}}=\frac{\partial \bar{L}}{\partial \gamma_{ab}}\Phi(\chi).
\end{equation}

Therefore, the new relation for the measure density $\Phi (\chi)$ leads to the usual theory. \\

Let's consider the special form of $\bar{L}$ that is going to be equal to $0$ by the constraint.

\begin{equation}
\bar{L} = \gamma^{ab}\partial_a X^{\mu} \partial_b X^{\nu} g_{\mu\nu} - \frac{\epsilon^{ab}F_{ab}}{\sqrt{-\gamma}},
\end{equation}

The modified action:

\begin{equation}
S = -\int d^2\sigma (\gamma^{ab}\partial_a X^{\mu} \partial_b X^{\nu} g_{\mu\nu} - \frac{\epsilon^{ab}F_{ab}}{\sqrt{-\gamma}}) \partial_h (\gamma^{hd} \sqrt{-\gamma} \partial_d \chi).
\end{equation}

Variations with respect to $\gamma_{ab}$, $X^{\mu}$ through the two-measure theory formalism restore the equations (\ref{eq:tab}), (\ref{eq:3}). Again variations with respect to $A_a$ recover the significant result for the tension being derived. \\

Now let us check the behavior of the equations under conformal transformations, $\Omega$:

\begin{equation}
\gamma_{hd} \rightarrow \Omega^2 \gamma_{hd}, \qquad \sqrt{-\gamma} \rightarrow \Omega^2 \sqrt{-\gamma}.
\end{equation}

Therefore, for the $\bar{L}$:

\begin{equation}
\bar{L} \rightarrow \Omega^{-2} \bar{L}.
\end{equation}

Then for the variation

\begin{equation}
\delta S = -\int d^2 \sigma \Omega^2 \bar{L} \Phi = 0.
\end{equation}

Then again we obtain the same result.

\begin{equation}
\bar{L} = 0.
\end{equation}

It means that the conformal symmetry is confirmed. \\

\section{Higher Dimensional Extended Objects (HDEO) without Galileon measure.}

The galileon symmetry is defined only in some special frame. In 2D the conformally flat frame is available. But there is no option to rigorously define a galileon symmetry in the higher dimensions. Only in 2D the modified $\Phi(\chi)$ could be simulteneously conformal invariant and possess the galileon symmetry. In the higher dimensions, then, there is no notion of galileon symmetry but the HDEO could possess conformal symmetry. Therefore, while generalizing our construction to the case of higher-dimensional branes we come to two possibilities different by their conformal behavior. Let's consider both cases. \\

I Conformally Invariant HDEO: \\

We could modify our measure $\Phi(\chi)$ to make it conformally invariant in any dimensions.

The generalized $\Phi(\chi)$:

\begin{equation}
\Phi(\chi) = \partial_h (\gamma^{hd} \sqrt{-\gamma} \partial_d \chi (-2\gamma^{ab} \partial_a \chi \partial_b \chi)^{\frac{d-2}{2}}).
\end{equation}

The same calculations as for the strings give us $\bar{L} = 0$ if $\bar{L}$ is homogeneous. Again we have the constraint on the gauge field. \\

Also, if $\bar{L}$ transforms homogeneously of degree $1$ in $\gamma^{ab}$, we may have

\begin{equation}
\Phi \bar{L} = \Phi (\gamma^{ab} \partial_a X^{\mu} \partial_b X^{\nu} g_{\mu\nu} + \sqrt{F_{ab} F^{ab}} + G^{ab} R_{ab} (\Gamma)),
\end{equation}

where $\Gamma_{ab}^{c}$ is independent of $\gamma_{ab}$ (1st order formalism). \\

In 4D we could use

\begin{equation}
\Phi = \epsilon^{abcd}F_{ab}F_{cd}.
\end{equation}

In 6D we could consider

\begin{equation}
\Phi = \epsilon^{\mu \nu \alpha \beta \gamma \delta} \epsilon_{abcd} \partial_{\mu} \varphi_a \partial_{\nu} \varphi_b \partial_{\alpha} \varphi_c \partial_{\beta} \varphi_d F_{\gamma \delta}.
\end{equation}

Moreover, we could consider the following measure:

\begin{equation}
\Phi = \partial_h(\gamma^{hd} \sqrt{-\gamma} \partial_d \chi (F_{ab}F^{ab})^{\frac{d-2}{4}}).
\end{equation}

Again we obtain the constraint $\bar{L} = 0$. Therefore, the internal structure of $\Phi$ does not reveal itself in the equations of motion. \\

II Non-Conformally Invariant HDEO: \\

If we do not care about the conformal invariance and take the same $\Phi (\chi)$ as in 2D, then we wouldn't get $\bar{L} = 0$. Instead we will have

\begin{equation}
\gamma_{ab} (-\frac{2}{\sqrt{-\gamma}} \frac{\partial \bar{L}}{\partial \gamma_{ab}}\Phi(\chi) -\partial^a \chi \partial^b \bar{L} - \partial^b \chi \partial^a \bar{L} + \gamma^{ab} \partial_e \chi \partial^e \bar{L}) \ne 0
\end{equation}

The constraint $\bar{L} = 0$ is not satisfied. Instead we get $\chi$ as a free dynamical degree of freedom. In the context of cosmology, this kind of models can provide interacting Dark Energy/Dark Matter scenarios \cite{f}. \\

\section{Gauging Matter in the Worldsheet with the Galileon Field}

Let us define the covariant derivative under the galileon gauge transformations:

\begin{equation}
D_a \varphi = \partial_a \varphi - ie(\partial_a \chi)\varphi.
\end{equation}

It is the example of minimally coupled galileon symmetry in the conformal gauge. \\

We consider "a gauge transformation" that introduces linear phase dependence in the complex scalar field:

\begin{equation} \label{eq:39}
\varphi \rightarrow \exp(ie\Lambda) \varphi,
\end{equation}

where $\Lambda = b_{a}\sigma^{a}$. \\

We see that it is covariant:

\begin{equation} \nonumber
D_a \varphi = \partial_a \varphi - ie(\partial_a \chi) \varphi = \partial_a(\varphi \exp({ie \Lambda})) - ie(\partial_a \chi)(\exp({ie\Lambda})) \varphi =
\end{equation}

\begin{equation} \nonumber
= \exp({ie\Lambda}) (\partial_a \varphi) + ie (\partial_a \Lambda) (\exp({ie\Lambda})) \varphi - ie (\partial_a \chi) (\exp({ie\Lambda})) \varphi -
\end{equation}

\begin{equation}
- i e b_a (\exp({ie\Lambda})) \varphi = \exp(ie\Lambda)D_a \varphi.
\end{equation}

Now we could add to the action a term of the form $\int\Phi(D_a \varphi)^{\ast} D^a \varphi d^2 \sigma$. The point is that the equations of motion won't change under the transformation (\ref{eq:39}).

\section{Conclusions}

In the present work, we have constructed a new measure of integration $\Phi(\chi)$ out of the galileon field $\chi$. $\Phi(\chi)$ may indeed be called a galileon measure because it is invariant under a galileon shift symmetry in the conformally flat frame. Previously, in the framework of the two measure theory the string tension was derived as a constant of integration, that is, it comes as an additional dynamical degree of freedom. Here a new $\Phi(\chi)$ gives the same result. Moreover, the degrees of freedom that appear at the action level do not appear in the equations of motion. Namely, the galileon higher derivative theory leads to the second order equations of motion.

When generalizing to higher-dimensional branes we are forced not to use the galileon symmetry for it is only available in 2D. We have considered several higher dimensional extended objects in which the internal structure of $\Phi(\chi)$ does not appear on the level of equations of motion. Having defined the galileon gauge transformations, we have gained the possibility to add an extra term to the action. Further developments in this area are needed and will be studied in a future publications.

\end{document}